\documentclass[11pt, a4paper]{article}

\usepackage[T2A]{fontenc}
\usepackage[utf8]{inputenc}
\usepackage[english]{babel}
\usepackage{amsmath}
\usepackage{graphicx}
\usepackage[margin=1in]{geometry}
\usepackage{authblk}
\usepackage{hyperref}

\newcommand{\orcid}[1]{\href{https://orcid.org/#1}{\textsuperscript{\textit{ORCID}}}}

\begin{document}

\title{Quantum Simulation of Si/SiGe Devices with Experimentally Calibrated Micromagnet Effects}

\author[1,*]{Andrii Sokolov}
\author[1]{Conor Power}
\author[2]{Mathieu Moras}
\author[2]{Claude Rohrbacher}
\author[1]{Brian Malone}
\author[3]{Sergey Amitonov}
\author[3]{Agostino Apr\`{a}}
\author[3]{Amir Sammak}
\author[3]{Nodar Samkharadze}
\author[1]{Elena Blokhina}

\affil[1]{Equal1 Laboratories Ireland, Dublin, Ireland}
\affil[2]{Equal1 Laboratories Canada, Sherbrooke, Queb\`{e}c, Canada}
\affil[3]{Equal1 Laboratories Netherlands, Delft, Netherlands}
\affil[*]{Author to whom any correspondence should be addressed. Email: \texttt{andrii.sokolov@equal1.com}}

\date{} 

\maketitle

\begin{abstract}
Silicon-based spin qubits in Si/SiGe heterostructures are a leading platform for scalable quantum computing, yet bridging the gap between theoretical computer-aided design (CAD) models and experimental reality remains a significant challenge. Standard simulations often fail to capture critical physical phenomena, such as interface dipoles, parasitic charge accumulation, and the magnetic hysteresis of on-chip micromagnets. In this work, we present a comprehensive, experimentally calibrated 3D simulation pipeline for a 6-dot Si/SiGe device. We refine the semiconductor band alignment and introduce a semi-empirical classical charge screening model to accurately capture the formation of parasitic wells and their suppression of gate lever-arms. Furthermore, we apply the Jiles-Atherton model to account for the hysteresis and pre-magnetization of integrated cobalt micromagnets, successfully reproducing the experimental resonant frequencies across all six qubits. By coupling these calibrated electrostatic and magnetic profiles into a time-dependent rotating wave approximation (RWA) Hamiltonian, we reproduce experimental observables, including microwave power chevrons. This framework provides a robust foundation for predicting device behavior, evaluating microwave line losses, and optimizing future scalable spin qubit architectures prior to fabrication.

\vspace{1em}
\end{abstract}

\section{Introduction}

Silicon-based spin qubits, particularly those defined in $\text{Si}/\text{SiGe}$ heterostructures, have emerged as a leading platform for scalable quantum computing. Their appeal lies in their exceptionally long spin coherence times, small physical footprint, and underlying compatibility with advanced semiconductor manufacturing processes. To transition from single- or two-qubit proof-of-concept experiments to larger, scalable quantum dot arrays (QDAs), accurate electromagnetic and quantum mechanical modelling of the device architectures is strictly required. However, bridging the gap between theoretical computer-aided design (CAD) models and experimental reality remains a significant challenge. Standard electrostatic solvers often rely on idealised material properties, such as Anderson's rule for band alignment, which fails to capture the interface dipoles and realistic band gaps present in fabricated $\text{Si}/\text{SiGe}$ devices. 

Furthermore, classical charge accumulation in buffer layers leads to parasitic quantum wells and screening effects that drastically alter the lever-arm coefficients of the control gates. In addition to electrostatics, the manipulation of individual spins via electric dipole spin resonance (EDSR) relies on local magnetic field gradients generated by integrated micromagnets. Predicting the amplitude and direction of the net vector magnetic fields is highly sensitive to the magnetization history and hysteresis of the magnetic materials, which are often overlooked in static finite element method (FEM) simulations. In this work, we present a comprehensive, experimentally calibrated 3D simulation pipeline for a 6-dot $\text{Si}/\text{SiGe}$ device equipped with bilateral charge sensors. We systematically address the discrepancies between standard numerical predictions and experimental data. Firstly, we refine the semiconductor band alignment and introduce a semi-empirical classical charge screening model to accurately capture the formation of parasitic wells and their influence on gate lever-arms. Secondly, we apply the Jiles-Atherton model to account for the hysteresis and pre-magnetization of the on-chip micromagnets, accurately reproducing the experimental resonant frequencies of the qubits. Finally, we couple our calibrated electrostatic and magnetic profiles into a time-dependent rotating wave approximation (RWA) Hamiltonian. By solving the system dynamics, we successfully reproduce experimental observables, such as microwave power chevrons, thereby validating our holistic modelling approach. This end-to-end framework provides a robust foundation for predicting device behaviour, evaluating microwave line losses, and optimising future scalable quantum dot architectures prior to fabrication.

\section{Device overview}

\begin{figure}[htb]
 \centering
\includegraphics[width=0.99\textwidth]{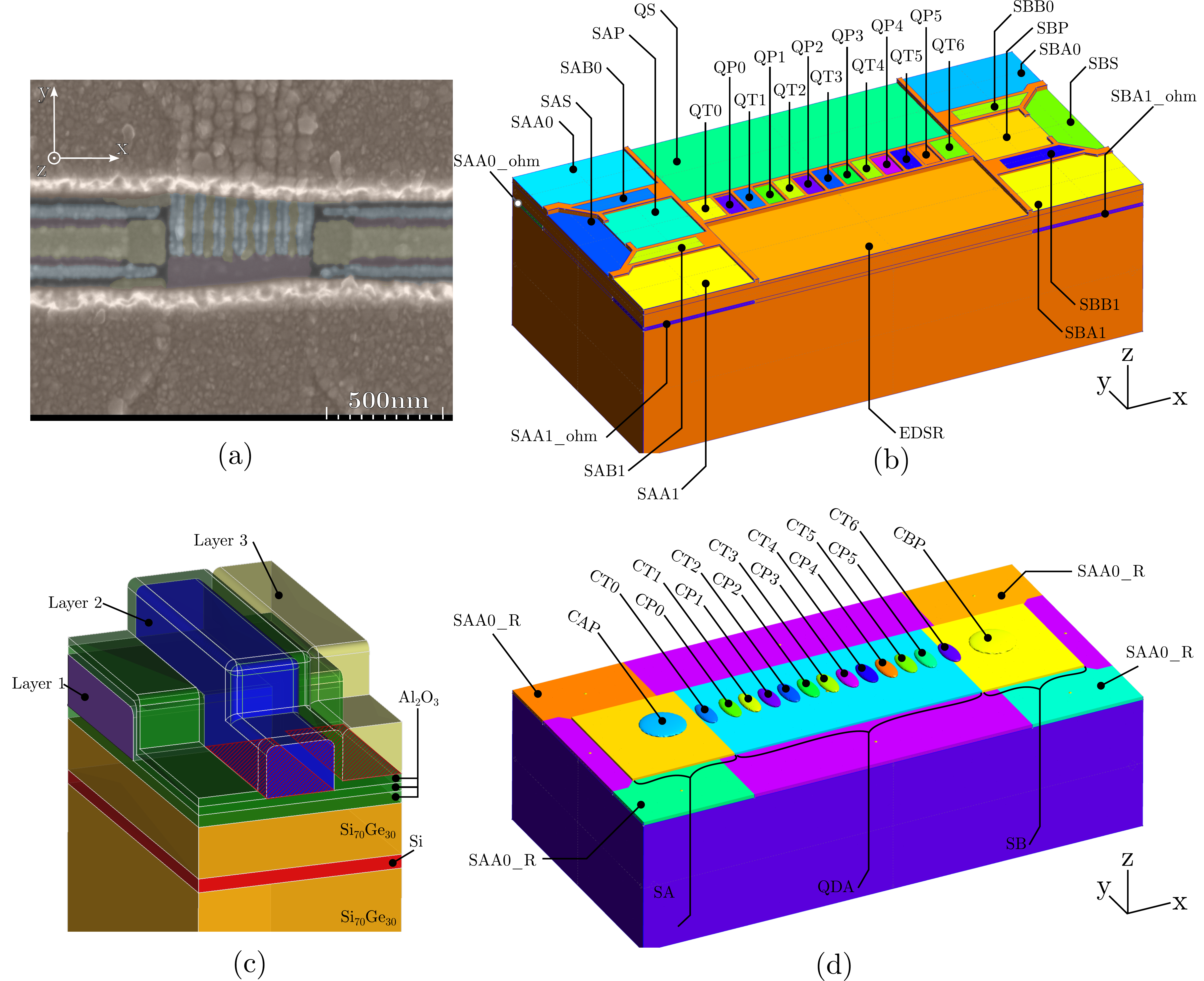}
 \caption{Finite element method model of the $\text{Si/SiGe}$ device under test. (a) SEM image of the device. (b) The 3D structure of the model with boundary conditions. (c) The material stack used in the model and the overlapping gates. (d) The same device with a hidden aluminium oxide layer. Here we show the QDA dot region, sensor dot regions, and reservoir domains. Additionally, parasitic charge regions are shown.}
\label{fig:layout}
\end{figure}

The simulated device is a $\text{Si/SiGe}$ heterostructure with three gate layers deposited on aluminium oxide (Fig.\,\ref{fig:layout}). It has two charge sensors on the sides of the 6-dot quantum dot array (QDA). The electromagnetic modelling of such a device requires taking into account many different aspects, which will be discussed in this section. 

The Finite Element Method requires building a convex 3D shape that represents the material composition of the device. The boundary conditions are usually defined by the gates deposited onto the structure. To construct our model, the in-plane ($XY$) structure is defined using the device layout and SEM images (Fig.,\ref{fig:layout}a), while the vertical ($Z$-axis) profile is deduced from the fabrication process specifications (Fig.,\ref{fig:layout}c). We are focused on the simulations of the quantum dots formed in the heterostructure; thus, we consider that the layer of $\text{Si}_{70}\text{Ge}_{30}$ is high enough (at least 200\,nm)  --- orange block in Figure\,\ref{fig:layout}\,c is covered by the $\text{Si}$ channel --- red block and then covered by the $\text{Si}_{70}\text{Ge}_{30}$ buffer --- orange block. Note that here we take into account neither the $\text{Si}_{x}\text{Ge}_{1-x}$ layer with decaying concentration of germanium nor bulk $\text{Si}$ wafer \cite{degliEsposti2024low}, since these layers do not affect the electromagnetic field in the Silicon channel. The $\text{Si}$-cap layer that is usually a 1\,nm height silicon covering the $\text{Si}_{70}\text{Ge}_{30}$ buffer layer is not included in the model, since it is not possible to create the mesh with such a fine layer, and the influence of this layer on the electromagnetic field in the dot will be minor. 

The $\text{Si}$ cap layer is covered by the aluminium oxide, and then three layers of gates are deposited on top of it. The layers are deposited one after another, each covered by the thin layer of the $\text{Al}_2\text{O}_3$. It is extremely difficult to create a 3D model of these gates (see Fig.\,\ref{fig:layout}\,c). It is also unnecessary because of the screening effects.  As a result, in this work, we are keeping only the non-overlapping parts of the gates and replacing them with the gate-boundaries \cite{philippopoulos2024analysis}. An example of this can be seen in Figure\,\ref{fig:layout}\,c where red shaded surfaces represent such a reduction. 

In Figure\,\ref{fig:layout}\,b we see all the boundary conditions used in the model. For simplicity, we group them in the following way: 
\begin{enumerate}
\item Sensor A has: 
\begin{itemize}
\item First layer screening gate $\text{SAS}$;
\item Second layer accumulation gates $\text{SAA0}$ and $\text{SAA1}$;
\item Second layer plunger gate $\text{SAP}$; 
\item Third layer barrier gates $\text{SAB0}$ and $\text{SAB1}$;
\item Side ohmic gates $\text{SAA0\_ohm}$ and $\text{SAA1\_ohm}$. 
\end{itemize}
\item Sensor B has:
\begin{itemize}
\item First layer screening gate $\text{SBS}$;
\item Second layer accumulation gates $\text{SBA0}$ and $\text{SBA1}$;
\item Second layer plunger gate $\text{SBP}$; 
\item Third layer barrier gates $\text{SBB0}$ and $\text{SBB1}$;
\item Side ohmic gates $\text{SBA0\_ohm}$ and $\text{SBA1\_ohm}$. 
\end{itemize}
\item Quantum dot array has: 
\begin{itemize}
\item The first layer screening gate $\text{QS}$;
\item The first layer EDSR line that for the electrostatic simulations serves as a screening gate as well $\text{EDSR}$;
\item The second layer plunger gates $\text{QP0}$, $\text{QP1}$, $\text{QP2}$, $\text{QP3}$, $\text{QP4}$, $\text{QP5}$;
\item The third layer tunnelling gates $\text{QT0}$, $\text{QT1}$, $\text{QT2}$, $\text{QT3}$, $\text{QT4}$, $\text{QT5}$, $\text{QT6}$;
\end{itemize}
\end{enumerate}

In addition to the layers that represent different materials, some artificial domains are included. There are so-called dot-regions where the classical charge transport is switched off. We include three separate dot-regions in our model --- $\text{SA}$ and $\text{SB}$ to analyse the quantum dots under the $\text{SAP}$ and $\text{SBP}$; and the $\text{QDA}$ to simulate the dots in the quantum dot array (Fig.\,\ref{fig:layout}\,d). All these domains include thin parts of the top and the bottom $\text{Si}_{70}\text{Ge}_{30}$ layers as well as the $\text{Si}$. 

The simulation of the accumulation effect in the heterostructure also requires the placement of specific reservoir domains $\text{SAA0\_R}$, $\text{SAA1\_R}$, $\text{SBA0\_R}$ and $\text{SBA1\_R}$ (Fig.\,\ref{fig:layout}\,d). We will discuss it in more detail in Section\,\ref{sec:acc_gate}.

Parasitic wells form in the buffer $\text{Si}_{70}\text{Ge}_{30}$ layer in a similar way to a SiMOS device. Usually, the parasitic wells are empty, since the mobility of charges in $\text{SiGe}$ is significantly lower than in $\text{Si}$ channel. However, if the voltage applied to the gates is high enough, there is a possibility of tunnelling between the main quantum dot and the parasitic one. This causes screening effects of the quantum dots, therefore, we included domains $\text{CAP}$, $\text{CBP}$ under plunger gates in the sensor, the domains $\text{CT0}$,\dots ,$\text{CT6}$ under tunnelling barriers and $\text{CP0}$,\dots ,$\text{CP5}$ under plunger gates. Manipulating the charge density in these domains allows us to take into account the effect of parasitic dots. This will be discussed in more detail in Section\,\ref{sec:finetuning}.

\section{Band alignment in Si/SiGe device and the influence of the frozen charges}\label{sec:band_al}

The band gap difference between $\text{Si}$ and $\text{Si}_{70}\text{Ge}_{30}$ provides a convenient way to confine electrons in the $Z$-dimension. As a result, the correct band alignment is a key element in device simulation. 

\begin{figure}[h!]
 \centering
\includegraphics[width=0.9\textwidth]{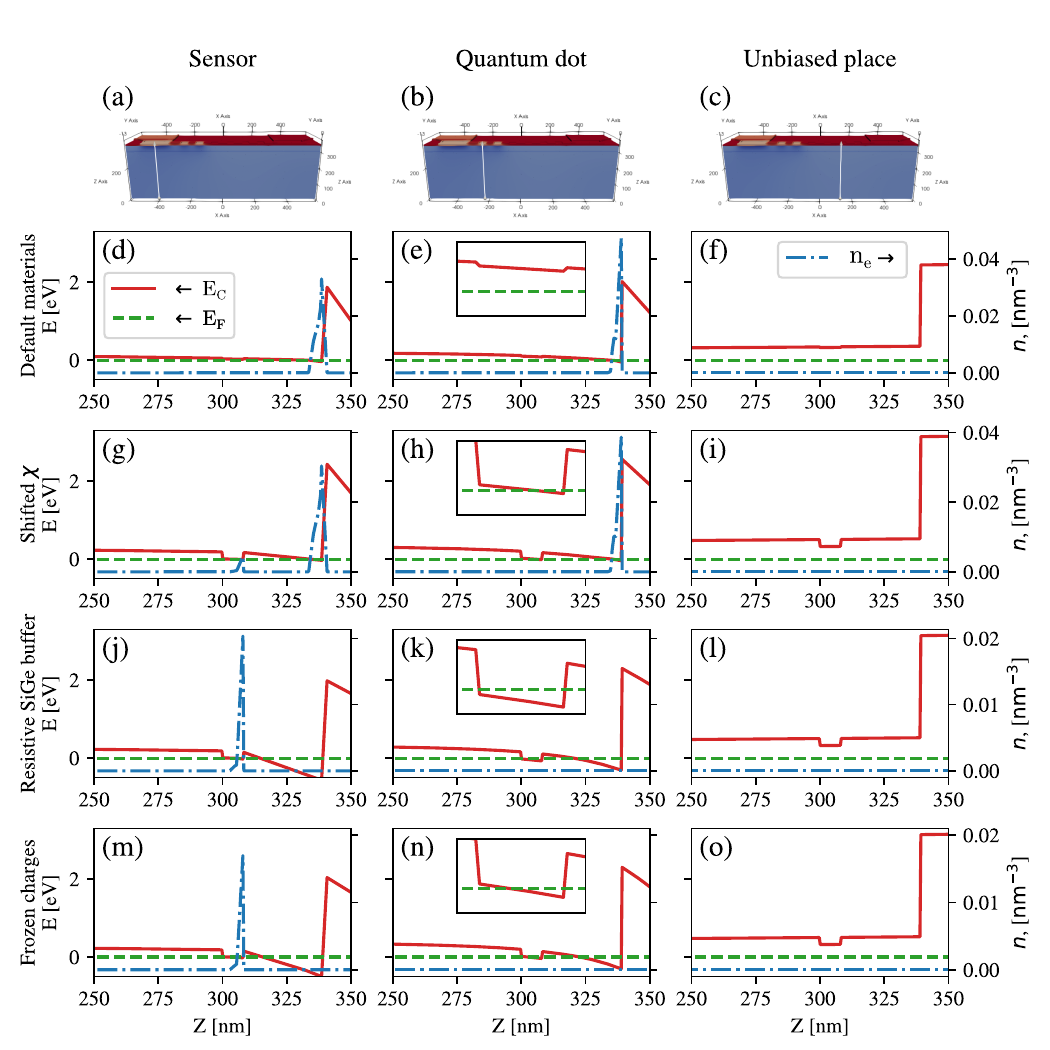}
 \caption{Examples of the transverse conduction band line-cuts at different compositions of materials and charges. The columns correspond to different line cuts: through the sensor, through one of the quantum dots, and through an unbiased region. Lines represent different material compositions: the default Anderson's rule; the case with shifted electron affinity $\chi$; the case where the $\text{SiGe}$ buffer layer is forced to be an insulator; the case that includes the influence of the semi-field parasitic quantum wells.}
\label{fig:band_alignment}
\end{figure}

The default approach in QTCAD$^\text{\textregistered}$ is using Anderson's rule for the band alignment\cite{beaudoin2022robust, prentki2023robust, philippopoulos2024analysis, sze1981physics}. Anderson’s rule states that the band alignment at a semiconductor heterojunction is determined by the difference in the electron affinities of materials and bandgap energies. Applying this approach to the simulated device, the model predicts the incorrect band gap between $\text{Si}$ and $\text{Si}_{70}\text{Ge}_{30}$, and between $\text{Si}$ and $\text{Al}_2\text{O}_3$ (see Fig.\,\ref{fig:band_alignment}\,d-f). 

In reality, electron affinities are influenced by surface charges and dipoles; however, in a heterojunction, the focus is on the interface between two materials, and there is no fundamental reason to expect that the dipoles at this interface will be the same as those found at a material-vacuum boundary. Therefore, the affinity of the $\text{Si}_{70}\text{Ge}_{30}$ and $\text{Al}_2\text{O}_3$ was shifted for the model to provide the correct band gaps (see Fig.\,\ref{fig:band_alignment}\,g-i). 

As one can see from Figures\,\ref{fig:band_alignment}\,g,h, if the classical charges are allowed in the $\text{Si}_{70}\text{Ge}_{30}$ layer, the Poisson solver fills the well with a classical charge until the conduction band edge is equal to the Fermi level at this point. Note that in Figure\,\ref{fig:band_alignment}\,h, $\text{Si}$ is in the so-called dot region where the classical charge is not allowed artificially to investigate the quantum properties of the dots. This effect causes the non-realistic screening of the gates. In reality, the accumulation gates are designed in such a way that the classical charges accumulate only in a $\text{Si}$ layer. The parasitic wells in the $\text{Si}_{70}\text{Ge}_{30}$ wells in the buffer get the frozen charges when they are tunnelling from the main wells from the $\text{Si}$ layer.  

Thus, the next optimisation step was to make the $\text{Si}_{70}\text{Ge}_{30}$ buffer an insulator. In this case, the classical charge is concentrated only in the $\text{Si}$ layer of the sensor, and the screening effect is absent (Fig.\,\ref{fig:band_alignment}\,j-l). 

However, as written before, there is some screening in the real device. This effect was modelled by adding the artificial negative charge density in the corresponding domains ($\text{CAP}$, $\text{CP0}$, $\text{CP1}$ in Figure\,\ref{fig:layout}\,d). The result of this is presented in Figure\,\ref{fig:band_alignment}\,m-o. More details on the fine-tuning are in Section\,\ref{sec:finetuning}.

\section{The influence of the classical charges}\label{sec:acc_gate}

The QTCAD$^\text{\textregistered}$ semiconductor solver is based on the classical semiconductor equation:
\begin{equation}
-\nabla\cdot\left(\varepsilon\nabla\varphi\right)=e\left(p-n+N_+-N_-\right)+\rho_0
\end{equation}
where $\varepsilon$ is the dielectric permittivity of the domain, $\varphi$ is the electric potential, $e$ is the elementary charge, $p$ and $n$ are the electron and holes densities, respectively, $N_+$ and $N_-$ are densities of donors and acceptors, and $\rho_0$ is a fixed background volume charge. 

Gates in the structure are represented by the gate boundary conditions:
\begin{equation}
\varphi = \varphi_\text{bias} - \frac{E_F}{e} + \varphi_F - \frac{E_w}{e},
\end{equation}
where $E_F$ is the Fermi level at the boundary nodes, $\varphi_F$ is the reference potential obtained after the band alignment, and $E_w$ is the gate metallic work function.  

In our simulations, we also have the ohmic boundary conditions
\begin{equation}
p-n+N_+-N_-=0.
\end{equation}
They are used to simulate the effect of the accumulation gates and reservoirs of electrons (boundaries $\text{SAA0\_ohm}$, $\text{SAA1\_ohm}$, $\text{SBA0\_ohm}$, and $\text{SBA1\_ohm}$ in Fig.\,\ref{fig:layout}\,b). 

\begin{figure}[htb]
 \centering
\includegraphics[width=0.95\textwidth]{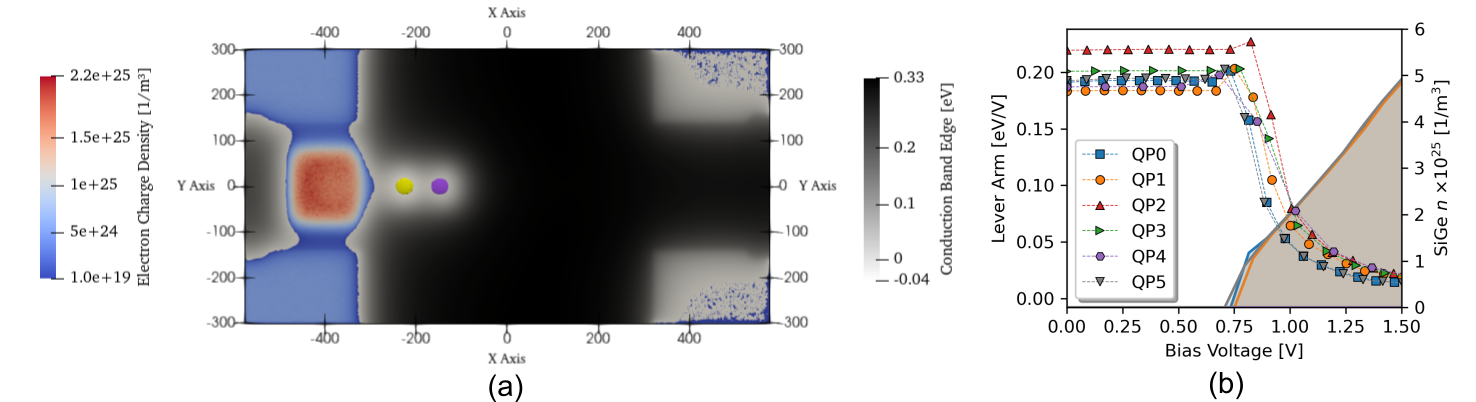}
 \caption{(a) The plane-cuts of the conduction band edge $E_C$ in the middle of the $\text{Si}$ channel and the electron charge density $n_e$ where it is non-zero. The ground states of the quantum dot under the $\text{QP0}$ --- yellow, and under the $\text{QP1}$ is violet. (b) Lever-arm variation with applied gate voltages and charges induced in $\text{SiGe}$ buffer layer.}
\label{fig:planecut}
\end{figure}

In reality, the accumulation gates serve as providers of the charges from the ion implantations to the dot array. However, the ion implantations are usually placed at $\sim 10$\,$\mu$m, which makes simulations of the accumulation gates very resource-intensive. In this work, we assume that the accumulation gates are already activated, so there are classical charges in the $\text{Si}$ channel under the accumulation gates. We emulate this effect by creating the specific reservoir domains ($\text{SAA0\_R}$, $\text{SAA1\_R}$, $\text{SBA0\_R}$, and $\text{SBA1\_R}$ in Fig.\,\ref{fig:layout}\,d) and adding the very weak $\sim10^{19}\,1/\text{m}^3$ n-doping to these areas. 

In Fig.\,\ref{fig:planecut}\,a is the result of the QTCAD$^\text{\textregistered}$ Poisson solver. The electron charge density is presented there as a blue-grey-red gradient. The left-hand side Sensor-A is switched on --- therefore, it is clear that there are relatively high classical charges under the accumulation gates. The Sensor-B on the right-hand side isn't switched on, and there are still visible classical charges there $\leq10^{19}\,1/\text{m}^3$ that should not be in reality, but these charges are away from the QDA, so they wouldn't influence the simulation. 

After the conduction band edge was calculated, the single-electron stationary Schr\"odinger equation was solved:
\begin{equation}
\left( -\frac{\hbar^2}{2m^*}\Delta + E_C(x,y,z) \right)\psi(x,y,z) = E\psi(x,y,z),
\end{equation}
where $\hbar$ is the reduced Planck constant, $m^*$ is the effective mass, $E_C(x,y,z)$ is the 3D conduction band edge, that is, the confinement potential in the case of electrons, $E$ is the eigenenergy, and $\psi(x,y,z)$ is the single-electron wavefunction. We designed the model in such a way that it is possible to solve the Schr\"odinger equation separately for both sensors and QDA (Fig.\,\ref{fig:layout}\,d). 

The sensitivity of the charge sensor depends on the sharpness of the Coulomb blockade peak; therefore, it is usually driven in a multielectron mode by applying a relatively high voltage to the $\text{SAP}$ gate. The integration over the sensor quantum dot gives $\sim665$ electrons in it.

\section{Fine-tuning of the Semiconductor model}\label{sec:finetuning}

Summarizing what was explained before, there are several parameters that allow our model to be fine-tuned. Firstly, the gates are made of palladium deposited on the thin layer of titanium. For the model, we assume the effect of the bottom surfaces of the gates; therefore, the biggest influence is made by the titanium work function, which is reported to be $\sim 4.3\,\text{eV}$ \cite{Michaelson1977}. However, this is hard to predict how much palladium is diffused into the titanium. Thus, the gate work function can be slightly corrected to a higher value to fine-tune the model, because the work function of the palladium is $\sim 5.6\,\text{eV}$\cite{Michaelson1977}. 

The charges in parasitic wells formed in the buffer $\text{SiGe}$ layer, explained in Section\,\ref{sec:band_al}, can also be used as a fine-tuning parameter, since it is not clear how much charge is actually in these parasitic wells. Here, we used the following approach: we calculated the number of electrons that can be in the parasitic wells if we make the $\text{SiGe}$ buffer with classical charge transport --- 251 electrons under the $\text{QP0}$, 300 electrons under $\text{QP1}$, and 3159 electrons under $\text{SAP}$; also, we define the size of the quantum well, just using the surfaces with charge density equal to $10^{19}$\,$1/\text{m}^3$. The quantum wells are shown in Fig.\,\ref{fig:layout}\,d. We place the uniform charge density in these dots, which have the same ratio as in the case of classical charges but have a lower number of electrons, assuming that the parasitic wells are partially filled. 

The classical charges in $\text{SiGe}$ buffer screen the electric field applied to the gate. This changes the lever-arm coefficients of the corresponding gates. This effect is easily observable experimentally and there are two clear pieces of evidence: 
\begin{itemize}
\item The bias-voltages applied to tunnel barriers and plunger gates in experiments are 20\% to 50\% higher than predicted voltages without screening.
\item There is a hysteresis in $C$$V$ curves explained by the charge accumulation in $\text{SiGe}$ buffer (close to $\text{Si}$-cap)\cite{10982705}.
\end{itemize}
To estimate the lever-arm variation due to the classical charge screening we allowed classical charge to be in $\text{SiGe}$ buffer. Then we selected points in $\text{Si}$ layer under corresponding gates and on top of the $\text{SiGe}$ buffer in $\text{Si}$-cap. The voltage seep for each gate gives the dependence $E_C(V_\text{gate})$. After that it is easy to calculate the derivative $\mathrm{d}E_C/\mathrm{d}V_\text{gate}$ that will indicate the lever-arm. Fig.\,\ref{fig:planecut}\,b presents the lever-arm of plunger gates (left axis) with voltage applied to the gate. It is easy to see that lever-arm significantly drops when charges begin to accumulate in $\text{SiGe}$ buffer. 

In reality, the plunger gates operate in the screened region when the barrier gates are in the unscreened region. Furthermore, this simulation captures the qualitative trend of the screening effect rather than predicting precise absolute values:
\begin{itemize}
\item The electrons in $\text{SiGe}$ buffer first tunnel from the reservoirs and then from one parasitic dot to another one, and there are merely tens of electrons, rather than hundreds. Therefore, to describe this accurately, one needs to apply the multielectron model.
\item The multielectron model requires a lot of resources, and still cannot catch the variations in $\text{Si}$-cap layer properties that dictate the variations of the $z$-shape of the potential. This effect is visible in the biasing scheme variation between left and right parts of the device, and between different devices.  
\end{itemize}

Therefore, in this work we use the semi-empirical model where the lever-arm transfer function is tuned according to the filling of the 1st electron to each dot in the array. 

\section{Simulations of Micromagnets}\label{sec:micromagnets}

Another important task is to be able to simulate the effect of the micromagnets. The micromagnets system for the device should be more and more complex for the scalable qubit architecture, therefore, the ability to predict the effect of micromagnets is extremely desirable. There are several known methods widely used in the field for the micromagnet simulations: 
\begin{itemize}
\item FEM calculation of the magnetic field created by the micromagnets with an empirically calculated constant magnetisation. This method is very fast, since the magnetisation considered to be constant and isn't re-calculated in each iteration. However, it doesn't show the non-full magnetisation of material and doesn't show how pre-magnetising of the micromagnets affect qubits. 
\item The Jiles-Atherton model is a model that on the one hand is FEM, so it does not have to be used in a separate framework. On the other hand, it catches the magnetisation-demagnetisation processes. \cite{539337}
\end{itemize}

The magnetic flux density $\mathbf{B}$ follows the relation
\begin{equation}
    \mathbf{B} = \mu_0\left( \mathbf{H} + \mathbf{M}\right),
\end{equation}
where $\mu_0$ is the magnetic constant, $\mathbf{H}$ is the magnetic field intensity and $\mathbf{M}$ is the magnetisation. In Jiles-Atherton approach, the effective $\mathbf{H}$-field $\mathbf{H}_e$ that acts on the magnetic moments within a magnetic domain is introduced. 
\begin{equation}
    \mathbf{H}_e = \mathbf{H} + \alpha\mathbf{M}.
\end{equation}
Here, $\alpha$ is the inter-domain magnetic coupling. The saturation of the magnetisation can be expressed by anhysteretic equation:
\begin{equation}
    \mathbf{M}_\text{an} = \mathbf{M}_\text{s}\left[ \coth\left(\frac{H_e}{a} - \frac{a}{H_e}\right) \right]\cdot\frac{\mathbf{H}_e}{H_e}
\end{equation}

Finally, the magnetisation as a function of the magnetizing field can be found from this differential equation:
\begin{equation}
    \mathrm{d}\mathbf{M} = \frac{1}{1+c} \cdot \frac{M_\text{an} - M}{\delta k - \alpha (M_\text{an} - M)} \mathrm{d}\mathbf{H} + \frac{c}{1+c}\mathrm{d}\mathbf{M}_\text{an}
\end{equation}
In summary, the Jiles-Atherton model has the following parameter for the tuning: 
\begin{itemize}
    \item $M_\text{s}$ is the saturation magnetisation in $A/m$;
    \item $a$ is the domain wall density in $A/m$;
    \item $\alpha$ is the dimensionless inter-domain magnetic coupling;
    \item $k$ is the pinning energy in $A/m$ that defines coercitivity;
    \item $c$ is the dimensionless reversibility;
    \item $\delta$ is 1 or -1 depending on the direction on change of the magnetisation. 
\end{itemize} 

\begin{figure}[h!]
 \centering
\includegraphics[width=0.99\textwidth]{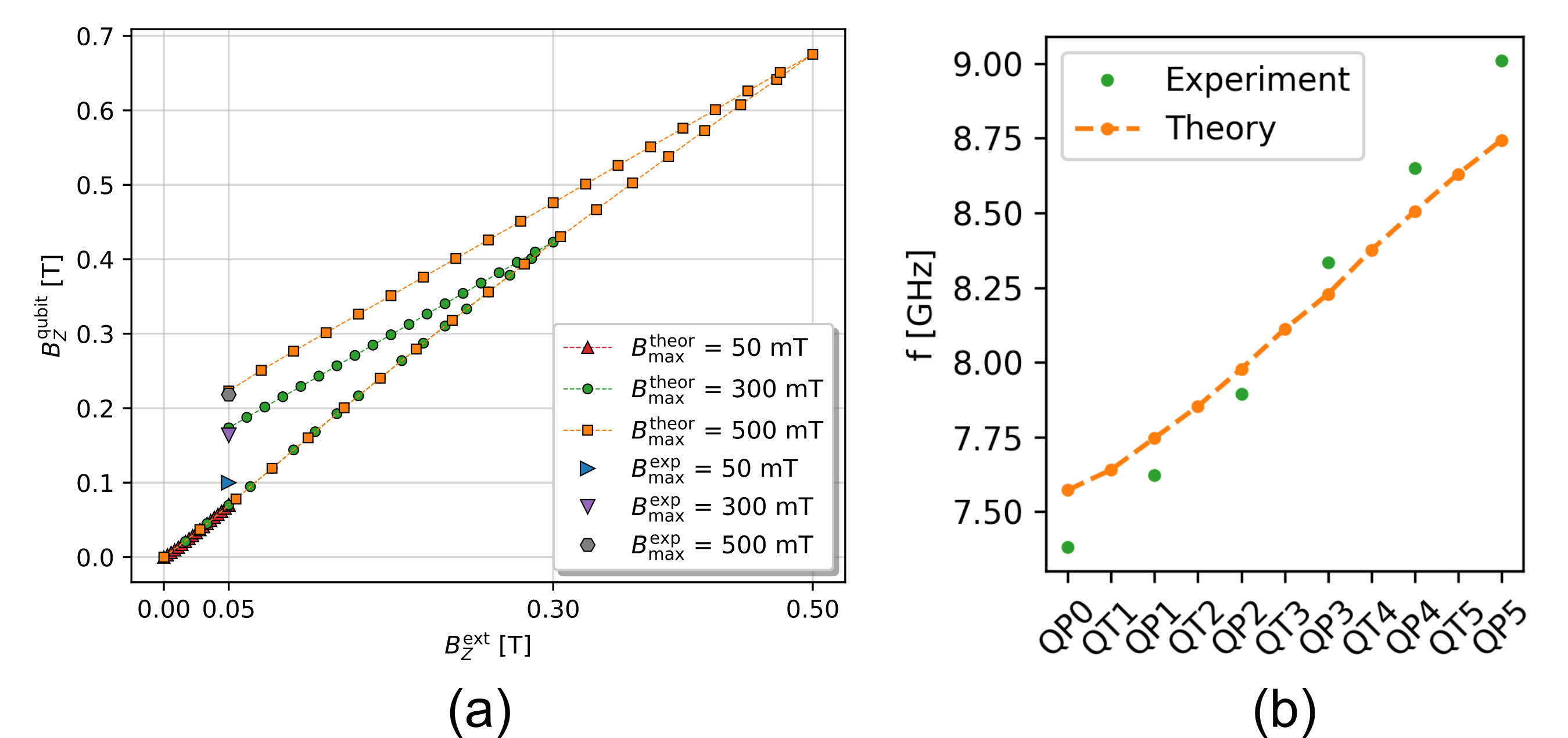}
 \caption{(a) Calibration step for the definition of Jiles-Atherton models parameters. (b) Experimental and simulated frequencies of all 6 qubits using the calibrated model. }
\label{fig:magnetic}
\end{figure}

This approach is used in the COMSOL Multiphysics simulation environment to compute the magnetisation and demagnetization effects and to calibrate these effects using the experimental data (see Fig.\,\ref{fig:magnetic}\,a). In simulations, we used the time-dependent approach when the external magnetic field was slowly ramped to the maximum (50\,mT, 300\,mT or 500\,mT), and then slowly ramped-down to the target field. In the calibration step, we fit the parameters of the Jiles-Atherton model described above to fit three known experimental dots (see Fig.\,\ref{fig:magnetic}\,a). The experimental field can be defined knowing the resonant frequency of the qubit: 
\begin{equation}
    f = \frac{g\mu_\text{B}}{h} B,
\end{equation}
where $g$ is the Land\'e g-factor that equals 2 for electrons, $\mu_\text{B}$ is the Bohr magneton, $h$ is the Planck constant and $B$ is the net magnetic field amplitude.

As observed, the experimentally derived magnetic field under a 50\,mT external field calibration—which effectively corresponds to the absence of a pre-magnetisation procedure --- is substantially higher than theoretical predictions. We attribute this discrepancy to the unknown initial magnetisation state of the micromagnets prior to the experiment. Nevertheless, when applied across the entire six-qubit array, the calibrated model yields resonant frequencies that are in excellent agreement with the experimental data (see Fig.,\ref{fig:magnetic}b). Although further fine-tuning of the model parameters is theoretically possible, the residual deviation between the simulation and the experimental results is now comparable in magnitude to the inherent device-to-device variation. These experimental uncertainties primarily stem from:
\begin{itemize}
\item Geometric variations introduced during lithography;
\item Fluctuations in the film thickness;
\item Inconsistencies in the deposition process and variations in the purity of the cobalt layer.
\end{itemize}

\section{Microwave Excitation of the Qubit}

After the pipeline for the semiconductor simulations in Sec.\ref{sec:band_al} - Sec.\ref{sec:finetuning} and micromagnets simulations in Sec.\ref{sec:micromagnets} it is possible to combine both of these simulations and use them as an input for the rotating-wave approximation (RWA)\cite{scully1997quantum} to compute the observable values. 

The RWA initially uses an unperturbed Hamiltonian:
\begin{equation}
H_0 = \sum_{n=0}^{N} \left( E_n - E_0 \right) |n\rangle \langle n|,
\end{equation}
where $E_n$ are eigenvalues of the system, $|n\rangle$ are the basis vectors (including spin states), $N$ is the total number of states. The RWA applies the time-dependent component as the perturbation:
\begin{equation}
    \Delta \tilde{V}_{mn} = V_{mn} - \delta_{mn}V_{00}.
\end{equation}

The excitation is considered to be time dependent --- sinusoidal with a given frequency $\omega_0$. 
\begin{equation}
    V(t) = \Delta \tilde{V} \cos (\omega_0 t).
\end{equation}
Here $\Delta \tilde{V}$ is the amplitude of the non-diagonal excitation due to the voltage applied to the microwave line. 

The full Hamiltonian is then: 
\begin{equation}
    H(t) = H_0 + \Delta \tilde{V} \cos (\omega_0 t). 
\end{equation}
The dynamics of the system represented by the Hamiltonian can be obtained by solving the corresponding high-level time-dependent Schr\"{o}dinger equation:
\begin{eqnarray}
    i\hbar \frac{\mathrm{d}}{\mathrm{d}t} | \psi(t) \rangle = H(t) | \psi(t)\rangle,\\
    |\psi(0)\rangle = |0\rangle. 
\end{eqnarray}
This is a well-known task that can be easily done by any of the packages for the high-level quantum simulations \cite{qutip5}. The high-level model obtained this way can have a probabilities to measure the states of electrons. 

\begin{figure}[h!]
 \centering
\includegraphics[width=0.95\textwidth]{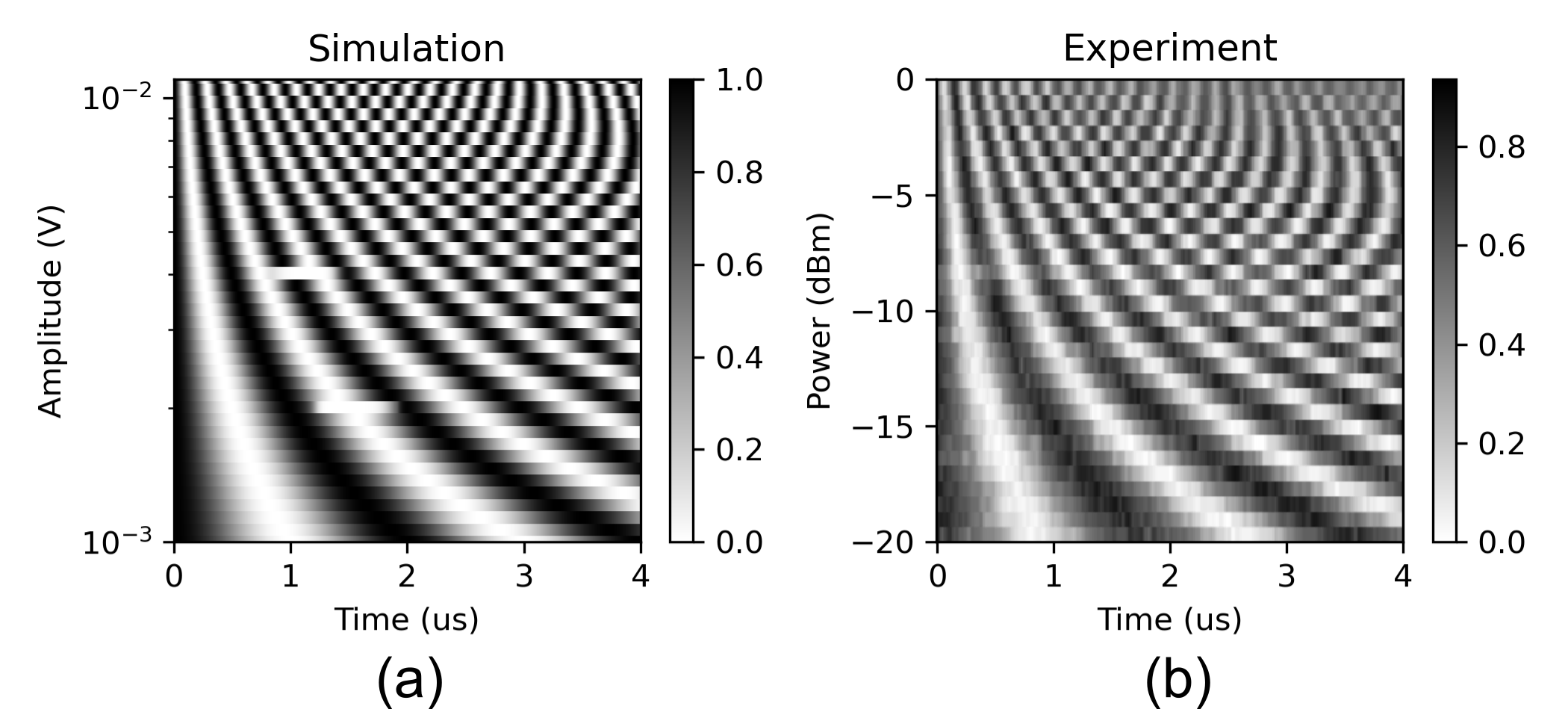}
 \caption{(a) Simulated power chevron of the device (b) Experimental power chevron of the device. }
\label{fig:chevron}
\end{figure}

The most obvious simulated experiment done by this technique is the so called power-chevron of the quantum dot when the Rabi oscillations have a different frequency compared with the power of the microwave excitation (Fig.\,\ref{fig:chevron}). It is clear that in simulated chevrons we cannot use the power, but have to use the voltage applied to the gate instead. However, it is possible to understand whether the simulated losses in the line correspond to losses predicted by other tools.

The signal is attenuated in the cryogenic line by 9\,dB, as a result we do not know exactly how much of the attenuation is in wire bonds (approximately 5 dB), the microwave line has losses of approximately 15 dB at the frequency of 8 GHz. With this in mind this, there is the following correspondence between voltage and power: 
\begin{itemize}
    \item 1\,mV of amplitude is -50\,dBm of power for 50\,$\Omega$, that is $\approx -49$\,dBm in experimental setup.
    \item 10\,mV of amplitude is -30\,dBm of power for 50\,$\Omega$, that is $\approx -29$\,dBm in experimental setup.
\end{itemize} 

\section{Conclusions}

In this work, we have developed and validated an end-to-end, experimentally calibrated simulation framework for a multi-dot $\text{Si}/\text{SiGe}$ heterostructure. By moving beyond idealised simulation parameters, we successfully captured the complex physical phenomena that dictate real-world device performance.

Our optimisation of the electrostatic model demonstrated that default band alignment theories, such as Anderson's rule, are insufficient for accurately describing the $\text{Si}/\text{SiGe}$ and insulator interfaces. By artificially shifting the electron affinities and modelling the $\text{Si}_{70}\text{Ge}_{30}$ buffer as an insulator augmented with discrete charge domains, we have effectively replicated the classical charge accumulation and parasitic well formation observed in experiments. This semi-empirical approach allowed us to accurately predict the suppression of plunger gate lever-arms caused by electrostatic screening, a critical factor for tuning multi-dot arrays.

Furthermore, we addressed the complexities of micromagnet design for EDSR by implementing the Jiles-Atherton model in our FEM simulations. By accounting for the hysteresis and unknown pre-magnetisation states of the cobalt micromagnets, our calibrated model achieved excellent agreement with the experimentally measured resonant frequencies across all six qubits. This highlights the necessity of tracking the dynamic magnetisation process rather than assuming constant material magnetisation.Finally, by coupling these static spatial simulations with a high-level time-dependent Schr"{o}dinger solver using the rotating wave approximation, we modelled the dynamic response of the qubit to microwave excitations. The resulting simulated power chevrons tightly matched the experimental Rabi oscillation data, allowing us to accurately estimate the real-world attenuation and signal loss in the cryogenic microwave lines. Ultimately, this calibrated simulation pipeline significantly bridges the gap between theoretical quantum CAD and experimental realization, providing a powerful tool for the design and optimization of next-generation scalable spin qubit processors.







\bibliographystyle{unsrt}

\bibliography{references.bib}

@book{sze1981physics,
  title     = {Physics of Semiconductor Devices},
  author    = {Sze, Simon M. and Ng, Kwok K.},
  year      = {1981},
  publisher = {John Wiley \& Sons},
  address   = {New York}
}

@article{degliEsposti2024low,
  title        = {Low disorder and high valley splitting in silicon},
  author       = {Degli Esposti, Davide and Stehouwer, Lucas E. A. and Gül, Önder and Samkharadze, Nodar and Déprez, Corentin and Meyer, Marcel and Meijer, Ilja N. and Tryputen, Larysa and Karwal, Saurabh and Botifoll, Marc and Arbiol, Jordi and Amitonov, Sergey V. and Vandersypen, Lieven M. K. and Sammak, Amir and Veldhorst, Menno and Scappucci, Giordano},
  journal      = {npj Quantum Information},
  volume       = {10},
  number       = {1},
  pages        = {32},
  year         = {2024},
  doi          = {10.1038/s41534-024-00826-9},
  note         = {Open access article}
}

@article{beaudoin2022robust,
  title={Robust technology computer-aided design of gated quantum dots at cryogenic temperature},
  author={Beaudoin, F. and others},
  journal={Applied physics letters},
  volume={120},
  number={26},
  year={2022},
  publisher={AIP Publishing}
}

@inproceedings{prentki2023robust,
  title={Robust Sub-Kelvin Simulations of Quantum Dot Charge Sensing},
  author={Prentki, R. and others},
  booktitle={2023 International Conference on Simulation of Semiconductor Processes and Devices (SISPAD)},
  pages={349--352},
  year={2023},
  organization={IEEE}
}

@article{philippopoulos2024analysis,
  title={Analysis and 3D TCAD simulations of single-qubit control in an industrially-compatible FD-SOI device},
  author={Philippopoulos, P. and others},
  journal={Solid-State Electronics},
  volume={215},
  pages={108883},
  year={2024},
  publisher={Elsevier}
}

@article{Michaelson1977,
  author    = {Herbert B. Michaelson},
  title     = {The work function of the elements and its periodicity},
  journal   = {Journal of Applied Physics},
  volume    = {48},
  number    = {11},
  pages     = {4729--4733},
  year      = {1977},
  doi       = {10.1063/1.323539}
}

@INPROCEEDINGS{10982705,
  author={Stampfl, F. and Godfrin, C. and Kubicek, S. and Baudot, S. and Raes, B. and De Greve, K. and Grill, A. and Waltl, M.},
  booktitle={2025 IEEE International Reliability Physics Symposium (IRPS)}, 
  title={{CV} Characterization of {Si/SiGe} Heterostructures at Cryo Temperatures}, 
  year={2025},
  volume={},
  number={},
  pages={1-5},
  doi={10.1109/IRPS48204.2025.10982705}}

@ARTICLE{539337,
  author={Bergqvist, A.J.},
  journal={IEEE Transactions on Magnetics}, 
  title={A simple vector generalization of the {Jiles-Atherton} model of hysteresis}, 
  year={1996},
  volume={32},
  number={5},
  pages={4213-4215},
  doi={10.1109/20.539337}}

@article{qutip5,
  title = {QuTiP 5: The Quantum Toolbox in {Python}},
  author = {
    Lambert, Neill and Gigu{`e}re, Eric and Menczel, Paul and Li, Boxi and
    Hopf, Patrick and Su{'a}rez, Gerardo and Gali, Marc and Lishman, Jake and
    Gadhvi, Rushiraj and Agarwal, Rochisha and Galicia, Asier and Shammah, Nathan and
    Nation, Paul and Johansson, J. R. and Ahmed, Shahnawaz and Cross, Simon and
    Pitchford, Alexander and Nori, Franco
  },
  journal = {Physics Reports},
  volume = {1153},
  pages = {1-62},
  year = {2026},
  issn = {0370-1573},
  doi = {10.1016/j.physrep.2025.10.001},
  url = {https://www.sciencedirect.com/science/article/pii/S0370157325002704},
}

@book{scully1997quantum,
  title     = {Quantum Optics},
  author    = {Scully, Marlan O. and Zubairy, M. Suhail},
  year      = {1997},
  publisher = {Cambridge University Press},
  address   = {Cambridge}
}

\end{document}